%% file: main.tex
\documentclass[%
 twocolumn,
 amsmath,amssymb,
 aps,
prb,
]{revtex4-2}

\usepackage{natbib,hyperref}

\usepackage{graphicx}
\usepackage{dcolumn}
\usepackage{bm}
\usepackage{mathtools}
\usepackage{amssymb}
\usepackage{amsmath}
\usepackage{amsthm}
\usepackage{graphicx}
\usepackage{bm}
\usepackage[dvipsnames]{xcolor}
\usepackage{epsfig}
\usepackage{amsfonts}
\usepackage{bbold}
\usepackage{multirow} 
\usepackage{footnote}
\usepackage{blkarray}
\usepackage[]{appendix}
\usepackage[makeroom]{cancel}
\makeatletter
\def\cantox@vector#1#2#3#4#5#6#7#8{%
  \dimen@.5\p@
  \setbox\z@\vbox{\boxmaxdepth.5\p@
   \hbox{\kern-1.2\p@\kern#1\dimen@$#7{#8}\m@th$}}%
  \ifx\canto@fil\hidewidth  \wd\z@\z@ \else \kern-#6\unitlength \fi
  \ooalign{%
    \canto@fil$\m@th \CancelColor
    \vcenter{\hbox{\dimen@#6\unitlength \kern\dimen@
      \multiply\dimen@#4\divide\dimen@#3 \vrule\@depth\dimen@\@width\z@
      \vector(#3,-#4){#5}%
    }}_{\raise-#2\dimen@\copy\z@\kern-\scriptspace}$%
    \canto@fil \cr
    \hfil \box\@tempboxa \kern\wd\z@ \hfil \cr}}
\def\bcancelto#1#2{\let\canto@vector\cantox@vector\cancelto{#1}{#2}}
\makeatother
\usepackage[]{hyperref}
\usepackage{soul} 

\newcommand{\expSTMESR}{~\cite{Baumann_Paul_science_2015,
Natterer_Yang_nature_2017,Choi_Paul_natnano_2017,Willke_Paul_sciadv_2018,Yang_Bae_prl_2017,
Willke_Bae_science_2018,Y_Bae_advanced_science_2018,Willke_Singha_nanolett_2019,
Willke_Yang_natphys_2019,Yang_Paul_prl_2019,yang_coherent_2019,Seifert_Kovarik_pr_2020,
Seifert_Kovarik_eabc_2020,
Weerdenburg_Steinbrecher_2020,Steinbrecher_Weerdenburg_2020,kim2021spin}}
\newcommand{\bias}{~\cite{Baumann_Paul_science_2015,
Natterer_Yang_nature_2017,Choi_Paul_natnano_2017,Willke_Paul_sciadv_2018,Yang_Bae_prl_2017,
Willke_Bae_science_2018,Y_Bae_advanced_science_2018,Willke_Singha_nanolett_2019,
Willke_Yang_natphys_2019,Yang_Paul_prl_2019,yang_coherent_2019}}

\begin{document}

\title{Many-body non-equilibrium effects in all-electric electron spin resonance} 

\author{Jose Reina-G{\'{a}}lvez}
\email{galvez.jose@qns.science}
\affiliation{Center for Quantum Nanoscience, Institute for Basic Science (IBS), Seoul 03760, Korea}
\affiliation{Ewha Womans University, Seoul 03760, Korea}
\author{Christoph Wolf}
\email{wolf.christoph@qns.science}
\affiliation{Center for Quantum Nanoscience, Institute for Basic Science (IBS), Seoul 03760, Korea}
\affiliation{Ewha Womans University, Seoul 03760, Korea}
\author{Nicol{\'a}s Lorente}
\email{nicolas.lorente@ehu.eus}
\affiliation{Centro de F{\'{i}}sica de Materiales
        CFM/MPC (CSIC-UPV/EHU),  20018 Donostia-San Sebasti\'an, Spain}
\affiliation{Donostia International Physics Center (DIPC),  20018 Donostia-San Sebasti\'an, Spain}

\begin{abstract}
Motivated by recent developments in measurements of electron spin resonances of individual atoms and molecules with the scanning tunneling microscope (ESR-STM), we study electron transport through an impurity under periodic driving as a function of the transport parameters in a model junction. 
The model consists of a single-orbital quantum impurity connected to two electrodes via  time-dependent hopping terms. The hopping terms are treated at the lowest order in perturbation theory to recover a Lindblad-like quantum master equation with electron transport.
 As in the experiment, the ESR-STM signal is given by the variation of the long-time DC current with the driving frequency. The density-matrix coherences play an important role in the evaluation of the ESR-STM signal. 
Electron correlation is included in our impurity mode. The charging energy $U$ has significant influence on the spin dynamics depending on the sign and magnitude of the applied DC bias. Our model allows direct insight into the origin of the ESR signal from the many-body dynamics of the impurity. 

\end{abstract}

\date{\today}

\maketitle

\section{Introduction}

The use of time-dependent techniques in the scanning tunneling microscope
(STM) at GHz frequencies ushered in  the acquisition of electron-spin
resonances (ESR) with the STM\expSTMESR.  These developments
grant access to new phenomena thanks to the unprecedented high-energy
resolution of ESR combined with the subatomic precision of the STM.
Examples are the analysis of elusive atomic configurations on
surfaces by measuring the actual magnetic moment of f-electron atoms
\cite{Natterer_Yang_nature_2017}, or the determination of isotopes of
single adsorbates \cite{Willke_Bae_science_2018}.

The ESR-STM technique measures the direct current (DC) through a localized spin impurity, such as single atoms or molecules, in the STM junction as the tip-sample bias is periodically modulated. At a certain
modulation frequency, the DC current experiences a variation that can
be detected. The ESR spectra are values of the junction DC current as
a function of driving frequency, typically in the GHz ($\mu$eV) range. To drive the localized electron spin, a suitable alternating electric field is fed either directly to the tip\bias\  or to the entire sample via an antenna \cite{Baumann_Paul_science_2015,Willke_Paul_sciadv_2018}. 
The mechanism that couples the electric field to a local magnetic moment is not clear
and substantial effort has been devoted to
try to understand under what circumstances ESR is produced (for a recent review please refer to Ref. [\onlinecite{Delgado_Lorente_pss_2021}]).
Clarifying the origin of ESR in the STM is not only conceptually, but also practically important, because the full development of the ESR-STM technique
requires a high degree of control to acquire meaningful signals.

In the present work, we address the effect of the transport
parameters in the ESR signal. Understanding how
transport affects the signal yields key information
on the way the ESR is produced. In previous publications \cite{J_Reina_Galvez_2019,J_Reina_Galvez_2021},
we have shown that a time-dependent modulation of the tunneling matrix elements
between electrodes and impurity is sufficient to produce a sizable ESR
signal. It is well-known that electric fields efficiently modulate
these transfer matrix elements \cite{Jauho_Wingreen_1994,arrachea-moskalets-2006}, and in turn, this modulation drives
the spin~\cite{J_Reina_Galvez_2019,J_Reina_Galvez_2021}.
Moreover, the suggested adiabatic motion of the impurity in
the time-dependent electric field \cite{Baumann_Paul_science_2015,Lado_Ferron_prb_2017}
would only increase the tunneling modulation.
Our model is based on a transport description of the electron
current in the presence of driving via the modulation of the tunneling matrix
elements. We treat the spin degrees of freedom via a reduced density
matrix, which allows us to develop a quantum master equation for the transport process under driving and with quantum spins~\cite{J_Reina_Galvez_2021}.

\begin{figure}
\includegraphics[width=1.\linewidth]{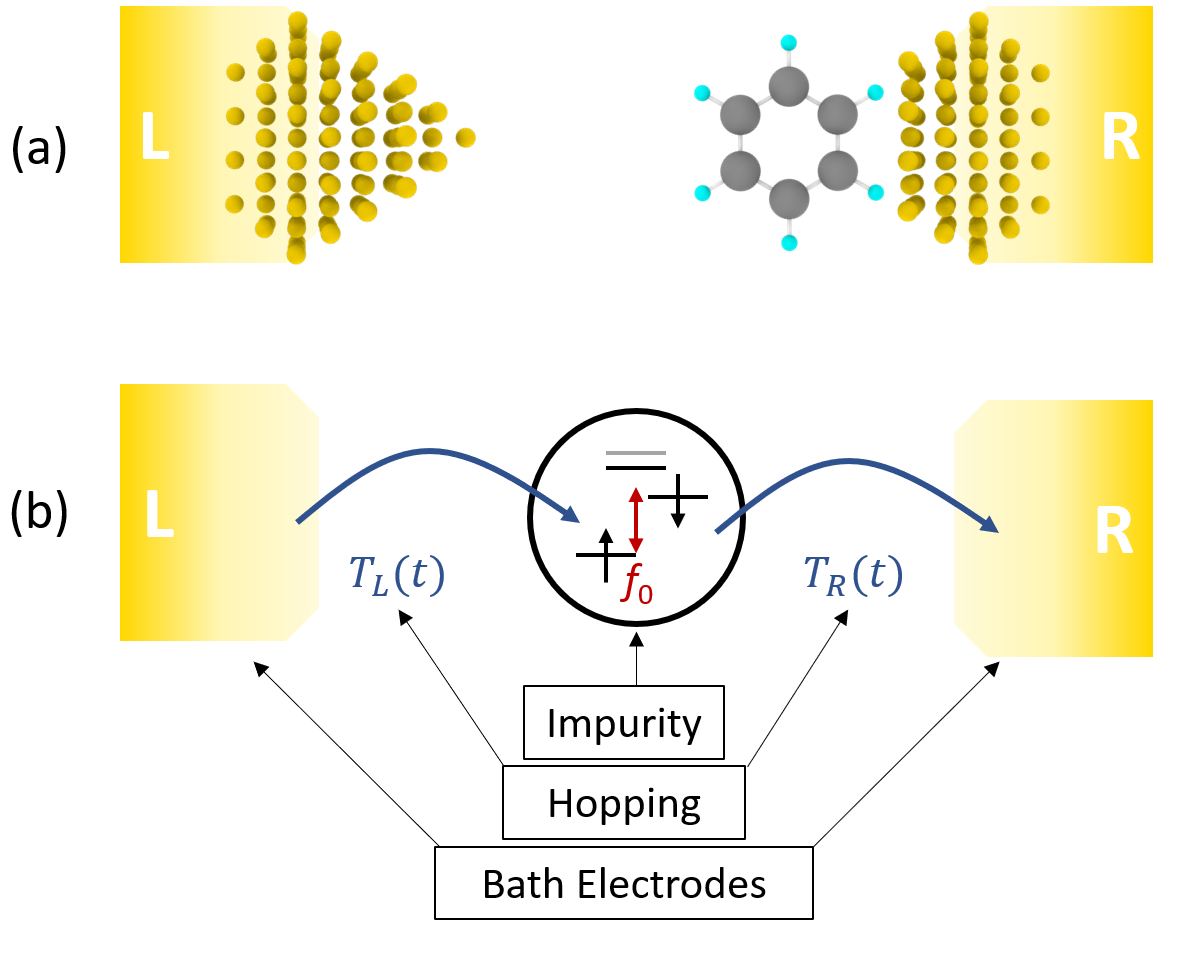}
\caption{(a) Scheme of the electron transport geometry for an impurity (here a  molecule) in a junction under an external drive given by the applied bias. In an ESR-STM setup, the left (L) and right (R) electrodes represent STM tip and substrate.
(b) The model used in our Hamiltonian representing the system in (a). A single orbital is connected via time-dependent hopping elements, $T_L(t)$ and $T_R(t)$, to the left
and right free-electron electrodes. Under an external magnetic field, the singly occupied spin up ($\uparrow$) and down ($\downarrow$) levels are split by
the Zeeman energy with Larmor or resonance frequency $f_0$. The orbital contains electron-electron correlation by the introduction of a charging
energy (or intra-orbital Coulomb repulsion) $U$.
}
\label{General_scheme}
\end{figure}

The article is organized as follows. In Sec.~\ref{sec_theory},
we summarize the model and the theoretical approach. In the present article, we put special emphasis
on clarifying the different equations and on how to treat the extended basis set to include electronic correlations under a finite charging energy $U$. We present the results of simulations with a set of parameters compatible with experimental ESR-STM setups in Sec.~\ref{Results}. The calculations explore
the behavior of the continuous wave (CW) ESR-STM signal (change in DC current as the driving frequency is changed) as a function of the DC bias. The results clearly show the role of the involved  states, the importance of having changing populations and coherences, as well as their influence
in the DC current that is ultimately the experimental observable. 
 The results corroborate the importance of coherent charge fluctuations to have a measurable signal in ESR-STM.

\section{Theoretical approach \label{sec_theory}}


Figure~\ref{General_scheme} shows a representative model of the type of system considered in this work. A central region that can be solved exactly is coupled via some hopping matrix elements to electron reservoirs. These hoppings need to be small compared to the typical energies of the central region in order to obtain a quantum master equation (QME) as will be shown in the following. We solve this QME in the long-time limit using Floquet's theorem that treats linear differential equations under a periodic drive~\cite{Grifoni_Hanggi_2021}. Finally in this section, we derive the equation for the time-dependent electronic current and for its DC component in this long-time limit.

\subsection{The model Hamiltonian}
The model for the full quantum system consists of a quantum impurity (a magnetic adsorbate, such as a single atom or molecule) tunnel-coupled to two electron reservoirs, see Fig. \ref{General_scheme}. The full system is described by
\begin{equation}\label{eq:ht}
H(t)=H_{\rm elec}+H_{\rm I}+H_{\rm T}(t),
\end{equation}
where the first term describes the two electrodes, the second term is
the impurity Hamiltonian and the third term is the tunneling Hamiltonian,
which is the only time-dependent one. The electrodes are
assumed to be described by one-electron states,
\begin{equation}\label{eq:hres}
H_{\rm elec}=\sum_{\alpha k\sigma }\varepsilon_{\alpha k}c^{\dagger}_{\alpha k \sigma }c_{\alpha k \sigma },
\end{equation}
 $\alpha$ identifies the electrode ($\alpha=$ L,R), while $\sigma =\uparrow, \downarrow$ is the electron spin projection along the quantization axis and $k$ is its momentum. Each electrode is characterized by a chemical potential $\mu_{\alpha}$ such that the total DC bias is $eV_{DC}=\mu_L-\mu_R$. Following Ref.~\cite{J_Reina_Galvez_2021}, the quantum impurity consists of a single orbital with intra-orbital
correlation represented by the charging energy $U$. The impurity is subjected to
 an external magnetic field such that its Hamiltonian is given by
\begin{eqnarray}\label{eq:himp0}
H_{\rm I}= \sum_{\sigma} \varepsilon d^{\dagger}_{\sigma }d_{\sigma }
+U \hat{n}_{d \uparrow} \hat{n}_{d\downarrow} +g\mu_B \mathbf{{B}}\cdot\mathbf{\hat{s}},
\end{eqnarray}
where $\varepsilon$ is the orbital energy of the impurity, 
$U$ is the corresponding Coulomb repulsion, and $\hat{n}_{d\sigma}=d^{\dagger}_{\sigma}d_{\sigma}$ is the occupation operator of the orbital.   
Its spin operator, $\mathbf{\hat{s}}$ has components
$\hat{s}^j=\hbar\sum_{\sigma,\sigma'}d^{\dagger}_{\sigma} \hat{\sigma}^j_{ \sigma \sigma'}d_{\sigma'}/2$, where $\hat{\sigma}^j$ ($j=x,y,z$) are the Pauli matrices. The last term of Eq. (\ref{eq:himp0}) is the Zeeman contribution to the Hamiltonian.

\begin{figure}
\centering 
\includegraphics[width=1.0\linewidth]{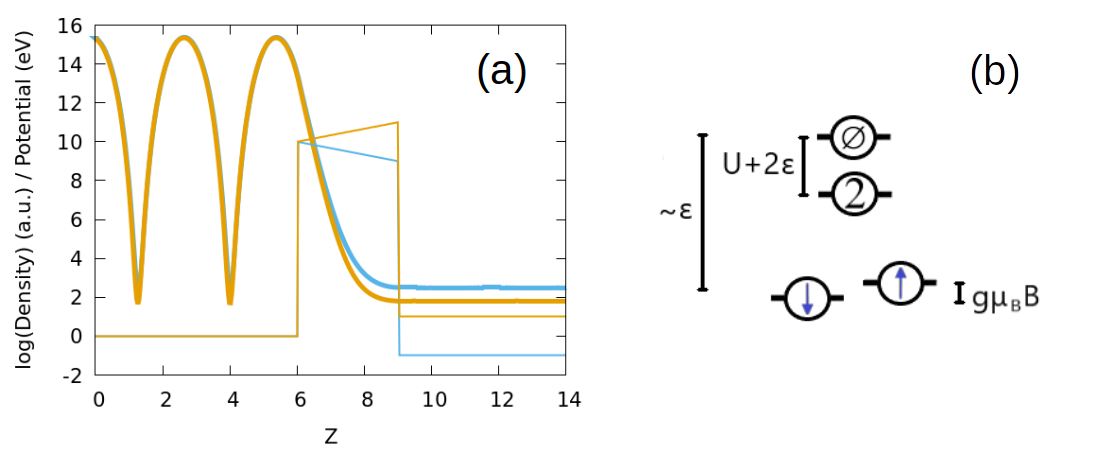}
\caption{(a) Barrier modulation represented here by two barriers at two different external electric fields. The transmitted wavefunction (given by the log of the density in thick orange and cyano curves) is much larger for one of the applied bias, illustrating the effect of the \textit{modulation of the hopping} that provides the electric-field coupling to the impurity spin. (b) Energy scheme of the quantum impurity. The four possible states are $|p\rangle=\uparrow, \downarrow, 2,\emptyset$. To simplify, we assume that the eigenstate basis $|l\rangle$ is, in first approximation, the same as the $|p\rangle$-basis. In order to be able to compare the energies of states with different number of electrons, we assume that the missing electrons are at the chemical potential of the electrodes at zero bias. The zero-electron state is  at zero energy.} 
\label{Energy_Sch}
\end{figure}

The coupling between the impurity and the two reservoirs is described by the tunneling Hamiltonian 
\begin{equation}\label{eq:ht0}
H_{T}(t)=\sum_{\alpha k\sigma}\left(T_{\alpha }(t)c^{\dagger}_{\alpha k\sigma }d_{\sigma}+T_{\alpha  }^*(t)d^{\dagger}_\sigma c_{\alpha k\sigma }\right).
\end{equation}
The periodic drive is introduced by a time-dependent hopping, $T_{\alpha} (t)$, parameterized as: \begin{equation}
T_{\alpha  }(t)=T_{\alpha }^0 \left[1+A_{\alpha }\cos (\omega t) \right], \label{hopping}
\end{equation}
following Refs.~\cite{J_Reina_Galvez_2019,J_Reina_Galvez_2021}. This approximation captures the effect of the driving electric field on the electron transfer probability because of the changing tunneling barrier. Figure~\ref{Energy_Sch} (a) shows a simple scheme for the modulation of the transmission of the wave function across one of the barriers, under varying external electric field. Although not needed, the presence of piezoelectric effects~\cite{Phark} would enhance the tunneling modulation in the time-dependent electric field.

Tunneling modulation is very efficient in driving the spin. In Ref.~\cite{J_Reina_Galvez_2019}, we showed that the tunneling modulation directly enters the Rabi flip-flop rate in an effective two-level system where electrons hop in and out the impurity. Indeed, the tunneling modulation implies an effective change of state of the impurity due to charge transfer from the electrodes.

To formulate the problem in terms of the reduced density matrix, we consider all possible configurations for zero, one and two electrons in the impurity. Figure~\ref{Energy_Sch} (b) shows an energy diagram with the four possible eigenstates of the simplest spin-1/2 system. These configurations are $|p\rangle $ with $p=\uparrow, \downarrow, 2,\emptyset$. The first two account for one electron states while the third and forth are labelling the spin singlets with two and zero electrons, respectively. The impurity Hamiltonian in this basis is given by:
\begin{eqnarray}\label{eq:himp}
H_{\rm I}&=&\sum_{p}\varepsilon_{p} |p\rangle \langle p|+\frac{g\mu_B}{2}\left\{ (B_x+iB_y)|\downarrow\rangle\langle\uparrow|\right.\nonumber \\ 
\crcr
&&+\left. (B_x-iB_y)|\uparrow\rangle \langle\downarrow| \right\}
\end{eqnarray}
where $\varepsilon_{p}$ takes the values $\varepsilon_{\sigma}  =\varepsilon +g\mu_B B_z\sigma$ for $\sigma=\uparrow$ or $\downarrow$, $\varepsilon_{2}=2\varepsilon+U$ and $\varepsilon_\emptyset=0$.
The tunneling Hamiltonian in this $|p\rangle$ basis set is:
\begin{equation}\label{ht}
H_T(t)=\sum_{\alpha k \sigma}\left(T_{\alpha }(t)c^{\dagger}_{\alpha k\sigma } |\emptyset\rangle \langle \sigma|+ T_{\alpha }(t) c^{\dagger}_{\alpha k\sigma } |\bar{\sigma}\rangle \langle 2|  + h.c. \right) 
\end{equation}
where $\bar{\sigma}$ indicates the opposite to the $\sigma$ spin projection. 

Since the impurity Hamiltonian does not depend on time, we can use the eigenbasis of the impurity to describe the reduced density matrix. This eigenstate basis is given by
\begin{equation}
H_{\rm I} |l\rangle = E_{l} |l \rangle.
\label{l-basis}
\end{equation}
From now on, Latin characters ($l,j,u,v,\dots$) refer to eigenstates that are combined electronic and spin configurations of the impurity. Accordingly, we write $H_T(t)$ in terms of the Hubbard operators $|l\rangle   \langle   j|$
obtained from these impurity many-body eigenstates~\cite{Hewson_book_1997},
\begin{equation*}
H_T(t)=\sum_{\alpha k \sigma lj} \left(T_{\alpha }(t)c^{\dagger}_{\alpha k\sigma } \lambda_{l j\sigma} |l\rangle   \langle   j|  + h.c. \right),
\end{equation*}
that explicitly contains the matrix element that reflects the change of the many-body configurations of the impurity $j$ of $N+1$ electrons to $l$ of $N$ electrons:
\begin{equation}
    \lambda_{lj\sigma} = \langle l | d_\sigma |j\rangle=\langle l|\emptyset\rangle \langle \sigma |j \rangle+\langle l|\bar{\sigma}\rangle \langle 2| j \rangle.
    \label{landa}
\end{equation}

\subsection{The quantum master equation}
We derive the QME by treating the coupling between the impurity and the reservoirs to the lowest  order in perturbation theory in $H_T$  like in Refs. [\onlinecite{schoen-94,koenig-96-1,koenig-96-2,spletts-06,Esposito_Galperin_prb_2009,cava-09,Bibek_Bhandari_2021_nonequilibrium,J_Reina_Galvez_2021}]. This approximation amounts to the Born-Markov approximation \cite{rammer_2007,Dorn_2021}).
The reduced density matrix in the impurity eigenstate basis set is 
\begin{equation}\label{rho}
\rho_{lj} (t) = \mbox{Tr}\left[\hat{\rho}_{T} (t) |l\rangle\langle j| \right],
\end{equation}
with  the trace taken over all the degrees of freedom of the total system and $\hat{\rho}_{T} (t)$ the time-dependent density matrix of also the total system~\cite{J_Reina_Galvez_2021,Bibek_Bhandari_2021_nonequilibrium}. 

The QME for $\rho_{lj}(t)$ is 
\begin{eqnarray}
\hbar\dot{\rho}_{lj}(t)&-& i\Delta_{lj}\rho_{lj}(t)= \sum_{vu}\left\{\left[\Gamma_{vl,ju}(t)+\Gamma^*_{uj,lv}(t)\right] \rho_{vu}(t) \right. \nonumber \\
&-&\left.   \Gamma_{jv,vu}(t)\rho_{lu}(t)-\Gamma^*_{lv,vu}(t)\rho_{uj}(t)\right\},
\label{rho_master_eq_6_0}
\end{eqnarray}
where we have denoted $\Delta_{l,j}=E_l-E_j$. All indices ($l,j,v,u$) refer only to many-body eigenstates of the impurity Hamiltonian, $H_{\rm I}$. 

The above QME, Eq. (\ref{rho_master_eq_6_0}), is physically meaningful in the limit of weak coupling between impurity and electrodes. Here, weak means that the induced broadening of the impurity levels is smaller than the typical separation between levels, $\Delta_{l,j}$. In this way, we make sure that the dynamics induced by the electrode is a small perturbation of the intrinsic impurity dynamics. In this limit, the different approaches to obtain a linear equation in the reduced density matrix yield the same QME \cite{Timm}.

The rates $\Gamma(t)$ can be written as the sum of two contributions per electrode $\alpha$: 
\begin{equation}
\Gamma_{vl,ju}(t) =\sum_\alpha\left[ \Gamma_{vl,ju,\alpha}^{-}(t) + \Gamma_{vl,ju,\alpha}^{+}(t)\right].
\label{rate}
\end{equation}
 These rates can be expressed as
\begin{eqnarray}
	\Gamma_{vl,ju,\alpha}^{-}(t)&=&\frac{i}{2\pi}\sum_{\sigma}  \lambda_{vl\sigma}\lambda_{uj\sigma}^*
       \left(1+A_{\alpha }\cos (\omega t) \right) \gamma_{\alpha\sigma}  \nonumber\\
       &\times& \int^\infty_{-\infty} d \epsilon f_\alpha (\epsilon)\left(
       \frac{1}{\epsilon-\Delta_{ju}+i\hbar/\tau_c}  \right.\nonumber \\ &+&  \left.e^{i\omega t} \frac{A_\alpha/2}{\epsilon-\Delta_{ju}\hbar\omega+i\hbar/\tau_c} + \right.\nonumber\\ &+& \left. e^{-i\omega t} \frac{A_\alpha/2}{\epsilon-\Delta_{ju}-\hbar\omega+i\hbar/\tau_c}  \right)
	\label{rateT}
\end{eqnarray}
and
\begin{eqnarray}
	\Gamma_{vl,ju,\alpha}^{+}(t)&=&-\frac{i}{2\pi}\sum_{\sigma}  \lambda_{lv\sigma}^*\lambda_{ju\sigma}
       \left(1+A_{\alpha }\cos (\omega t) \right) \gamma_{\alpha\sigma}  \nonumber\\ &\times& \int^\infty_{-\infty} d \epsilon (1-f_\alpha (\epsilon))\left( \frac{1}{\epsilon+\Delta_{ju}-i\hbar/\tau_c}  \right.\nonumber \\ &+& e^{i\omega t} \frac{A_\alpha/2}{\epsilon+\Delta_{ju}+\hbar\omega-i\hbar/\tau_c}  \nonumber \\ &+& \left. e^{-i\omega t} \frac{A_\alpha/2}{\epsilon+\Delta_{ju}-\hbar\omega-i\hbar/\tau_c}  \right).
	\label{rateTle}
\end{eqnarray}
The Fermi occupation function is given by $f_\alpha(\epsilon)=1/\left(e^{\beta_\alpha(\epsilon-\mu_{\alpha})}+1\right)$  where $\beta_\alpha$ is the inverse temperature times the Boltzmann constant for electrode $\alpha$. Additionally, $\gamma_{\alpha\sigma}$ is the level broadening due to the hopping, $T^0_\alpha$, to electrode $\alpha$ for spin $\sigma$:
\begin{equation}\label{pol}
    \gamma_{\alpha\sigma}=2\pi D_{\alpha\sigma}|T^0_\alpha|^2,
\end{equation}
that depends on the spin-dependent density of states, given by 
\begin{equation}
D_{\alpha\sigma}=D_{\alpha}(1/2+\sigma P_\alpha).
\label{Pol}
\end{equation}
Here, $\sigma=\pm1/2$ and $P_\alpha$ is a real number between -1 and 1 giving the spin polarization of the electrode, and  $D_\alpha$ is
the density of states of the electrode at the Fermi energy.

The above rate expressions neglect the time-dependence of the electrodes \cite{J_Reina_Galvez_2021}. This approximation is valid when the AC amplitude is much smaller than the DC component of the bias, otherwise Eq. (\ref{rateT}) should include further Bessel functions to take into account the time-dependence of the electrode's Green's function~\cite{Jauho_Wingreen_1994,arrachea-moskalets-2006}. 

The physical interpretation of the rates is straightforward. The rates are proportional to $\gamma_{\alpha\sigma}$, Eq. (\ref{pol}), that is the usual broadening induced
by the hopping-matrix elements and the density of states of the electrodes. Whether the process involves electrons or holes is contained in the appearance of the Fermi occupation factors. The expressions given in Eqs. (\ref{rateT}) and (\ref{rateTle}) contain the $\lambda$ matrix elements that take in the right weights of each impurity state. Finally, the factors including the magnitude of the hopping modulation $A_\alpha$ take into account whether the electron-transfer process involves the absorption or emission of a photon from the microwave field. 

Finally, a finite $1/\tau_c$ improves the convergence of the Green's function. Its inclusion leads to a small renormalization (or Lamb shift) of the spectrum.

\subsection{The long-time limit}
The rate is periodic in time at a fixed drive of frequency $\omega/2 \pi$ and can be expanded in terms of Fourier components allowing us to 
express all equations in Floquet components. We introduce the Floquet index $n$ as the Fourier index of the rate~\cite{Grifoni_Hanggi_2021}:
\begin{eqnarray}
\Gamma_{vl,ju,\alpha}(t) &=&\sum_{n} e^{-in\omega t}\;\Gamma_{vl,ju,\alpha;n}(\omega).
\label{Rate_Floquet_form}
\end{eqnarray}

From Eqs. (\ref{rho_master_eq_6_0}) and (\ref{Rate_Floquet_form}), we can write the Floquet master equation,
\begin{eqnarray}
 \Delta_{lj}\rho_{lj;n}  &+& n\hbar \omega  \rho_{lj;n}= \nonumber \\
 i\sum_{vu; n'}\{[\Gamma_{vl,ju;n'}(\omega)&+&{\Gamma}_{uj,lv;-n'}^{*}(\omega)]\rho_{vu;n-n'} \nonumber \\
 -{\Gamma}_{lv,vu;-n'}^{*}(\omega)\rho_{uj;n-n'}&-& \Gamma_{jv,vu;n'}(\omega)\rho_{lu;n-n'}\}.
\label{rho_master_eq_Floquet}
\end{eqnarray}

\subsection{Expressions for the electronic current}
 
The  current flowing out of electrode $\alpha$ is defined as $I_{\alpha}=-e\frac{d\langle N_{\alpha}\rangle}{dt} $. This translates
into the usual Meir-Wingreen formula \cite{meir_wingreen_prl_1992}, where
now the matrix elements of all quantities appear in terms of  many-body eigenstates, $l,j,u$ (see Ref.~\cite{J_Reina_Galvez_2021}):
\begin{equation}
I_{\alpha}(t)=
\frac{2e}{\hbar}\sum_{lju} \mbox{Re} \left\{\rho_{lu}(t) \left[ \Gamma_{lj,ju,\alpha}^{-}(t) - \Gamma_{lj,ju,\alpha}^{+}(t)\right] \right\}.
\label{current_t}
\end{equation}
Using $I_L=-I_R$, we symmetrize the current by making $I=(I_{L}+I_L)/2=(I_{L}-I_R)/2$ and the above expression can be rewritten as
\begin{eqnarray}
I(t)=
-\frac{2e}{\hbar}\sum_{lju} \mbox{Re}&& \left\{\rho_{lu}(t) \left[ \Gamma_{lj,ju,R}^{-}(t) + \Gamma_{lj,ju,L}^{+}(t) -\right.\right. \nonumber \\ 
&&\left.\left. \Gamma_{lj,ju,L}^{-}(t)-\Gamma_{lj,ju,R}^{+}(t) \right] \right\}.
\label{current_sym}
\end{eqnarray}
This expression differs from previous approaches  because it now contains the contribution of the coherences of the density matrix not only the populations \cite{Rossier_prl_2009,Delgado_Rossier_prb_2010,Yinan_Choi_2021}. 
We will show that under certain conditions the coherences are crucial for the correct calculation of the ESR signal.

Since CW ESR-STM experiments measure the DC current in the long-time limit, we express it in Floquet components as:

\begin{eqnarray}
I(\omega)=
-\frac{2e}{\hbar}&&\sum_{lju;n'} \mbox{Re}  \bigg\{\rho_{lu;-n'}(\omega) \times \nonumber\\  && \left[ \Gamma_{lj,ju,R;n'}^{-} + \Gamma_{lj,ju,L;n'}^{+} -\right. \nonumber \\ 
&&\left. \Gamma_{lj,ju,L;n'}^{-}-\Gamma_{lj,ju,R;n'}^{+} \right]  \bigg\}.
\label{Current_floquet_simp}
\end{eqnarray}

Finally, let us emphasize that the full description based on a QME
is possible when keeping to the lowest order in the hopping terms. This order is
sufficient when the impurity level lies within the two Fermi levels. However, outside this
bias window, higher-order terms may become comparable to or larger than
the lower-order term. These higher-order terms contain sums over intermediate
states opening the possibility to co-tunneling processes 
and to Kondo scattering (see for example Ref. \cite{J_Reina_Galvez_2019} and \cite{Korytar2011,DJ2017}). In the present approach these processes are absent.

\input{Resultados_2}

\section{Summary and conclusions}

This work explores the ESR signal in the DC current through a quantum impurity connected to two electron reservoirs under bias. The model is intended to reproduce the conditions of ESR-STM, where the applied bias contains a DC component and a AC component usually in the GHz frequency range. We extend previous work~\cite{J_Reina_Galvez_2021} to include finite intra-atomic correlation, and we show its impact on the DC-bias dependence of the ESR signal. Our theory is based on a Linblad-like QME that was obtained by keeping the modulation of the tunneling matrix element to lowest order. This limits the transport regime to the sequential or on-resonance one. This situation seems to be similar to transition-metal impurities, molecules, or alkali metal dimers that have s-electrons close to the Fermi energy of the substrate~\cite{Jinkyung_Hyperfine_2022, zhang_electron_2022, kovarik_electron_2022,Kawaguchi_2022}. We only treat spin-1/2 systems in the present study but the extension to larger spin systems can be achieved with relative ease.

The addition of the impurity charging energy, $U$, breaks the electron-hole symmetry of the system. This has wide ranging implications for the transport when taking into account the opening and closing of different transport channels as the applied DC bias varies.
As a consequence, spin-1/2 systems such as the ones of Refs.~\cite{Jinkyung_Hyperfine_2022, zhang_electron_2022, kovarik_electron_2022,Kawaguchi_2022} should exhibit a bias-sign dependence of the ESR signal in the experiment. 



Our work highlights the importance of properly including the complete reduced density matrix in the calculation of the ESR signal. In the open-channel case, we found that the ESR signal is proportional to the coherences or off-diagonal elements of the density-matrix. However, in the closed-channel region, the diagonal elements or populations play a significant role. Therefore, the ESR-induced change in the DC current can be indicative of coherences or population changes of the system depending on the transport regime. 
 
 The present theory is based on a charge-fluctuation description where the impurity charge is changing during the electron transport process, and the fluctuations induce the spin-flip processes that in turn lead to the  ESR-signal as long as driving and polarization are maintained. Our results emphasize  the need to correctly treat the coherent charge fluctuation and include the coherence in the description of the full transport processes, not only for the evaluation of the impurity's population but also in the equation of the electron current.

\begin{acknowledgments}
We are pleased to thank our collaborators for important discussions. A non-exhaustive list of the
many contributors to our discussions is:
L. Arrachea,
D.-J. Choi,
F. Delgado,
F. Donati,
J.-P. Gauyacq,
A. J. Heinrich,
S.-H. Phark.\\
\newline
This work was supported by the Institute for Basic Science (IBS-R027-D1). Further financial support from projects RTI2018-097895-B-C44 and PID2021-127917NB-I00 funded by MCIN/AEI/10.13039/501100011033 is gratefully acknowledged. Funded by the European Union. Views and opinions expressed are however those of the author(s) only and do not necessarily reflect those of the European Union. Neither the European Union nor the granting authority can be held responsible for them. 
\end{acknowledgments}

\bibliographystyle{apsrev}
\bibliography{references.bib}

\end{document}

%% file: Resultados_2.tex
\section{Results}
\label{Results}

Our model consists of a $S=1/2$ impurity that is weakly connected to two electrodes under a finite DC bias and a CW drive.  
Our aim is to explore the behavior of the ESR signal as the DC voltage is varied for a set of parameters intended to mimic conditions found in ESR-STM experiments. 
\subsection{Model parameters}
\label{parameters}

The model parameters are chosen under the proviso of obtaining a strong ESR signal of a $S=1/2$ system weakly connected to two electrodes under electrical driving. To achieve this, we need:

\begin{enumerate}
    \item an imbalance in the transport-electron spin in order to make the main rates different from zero. This is achieved by having different spin-polarization of the electrodes.
    
    \item a predominant long-time average population of one electron in the impurity, otherwise the system does not behave like a $S=1/2$. 
    
    \item an electronic level, $\varepsilon$, within the DC-bias range. 
    
    \item to flip the transport spin using a magnetic field transversal to the electron spin polarization.
    
    \item a modulation of the tunneling hopping with the spin-polarized electrode by the oscillating electric field. 

    \item low temperature. We take 1 K for both electrodes.
\end{enumerate}

In our calculations, we achieve the above conditions with the following parameters: 1. The  left electrode has a polarization of $P_L= 0.45$ in Eq. (\ref{Pol}). Increasing the polarization up to 100\%  will increase the ESR signal amplitude. 2. To stabilize the charge state, we apply different couplings with $\gamma_{R}=20\times\gamma_{L}=5\ \mu$eV. This coupling asymmetry is often found in experiments, where the impurity couples more strongly to the substrate than the STM tip. The DC bias drop is $eV_{DC}=\mu_L-\mu_R$. We use the model of a double-barrier tunnel unction \cite{Tu_X_double-barrier} and assume an asymmetric DC bias drop where $\mu_L=(1-\eta)eV_{DC}$ and $\mu_R=-\eta eV_{DC}$ with the factor $\eta=\gamma_L/(\gamma_L+\gamma_R)=1/21$. This means that the bias drop takes places mostly on the left electrode. 3. The energy of our model is set by $\varepsilon=-10$ meV. In addition, the electronic states are assumed to have an intrinsic width of $\hbar/\tau_c=10 \;\mu$eV. 
In order to explore the interplay of the many-body states in ESR processes, we take a fixed charging energy close to the electronic level energy, of $U=3|\varepsilon|/2=15$ meV. 4. In order to flip the spin, defined along the $z$-axis of the spin polarization, $P_L$, we apply a $B$-field component along the $x$-axis perpendicular to the $z$-axis component. The magnetic field is taken as $\mathbf{B}=(0.6, 0, 0.1)$ T, which gives a Larmor frequency of approximately 17 GHz. The largest ESR signal takes place for a magnetic field completely aligned with the $x$-axis in good agreement with experiments~\cite{Rotation}. 5. The modulation of the tunneling matrix element is $A_L=50\%$, Eq. (\ref{hopping}) and applied only to the left electrode, which is the polarized one. Since the right electrode is not spin polarized, $A_R$ does not contribute to the resonance, but only to the background current.


\subsection{Non-zero rates: the opening of transport channels with applied bias}

A transport channel opens when the corresponding rates, Eq. (\ref{rate}) are different from zero. Inspection of Eq. (\ref{rateT}) shows that this occurs when two conditions are met: The first one is energy conservation, largely controlled by the Fermi factors. The energy conservation implies that the change of state has to be compensated by the applied bias. Under our present conditions, the bias drop takes place largely at the left electrode, then $\Delta_{v,l}=E_v-E_l$ has to be larger than $\mu_L=eV_L=(1-\eta)eV_{DC}$. This is due to the appearance of a term $f(\Delta_{v,l})$ in Eq. (\ref{rateT}) when $1/\tau_c\rightarrow 0^+$. 
The second condition is that the sequential transport process leads to a change in the charge state of the impurity such that  $\lambda_{vl\sigma}\neq 0$ when $v$ and $l$ differ in one electron of spin $\sigma$. Then, the difference in energy $\Delta_{v,l}$ in the rates Eq. (\ref{rate}) always addresses states differing by one electron.

\subsection{DC-bias dependence of the ESR signal}

The DC-bias will determine when the transport channels of the system opens. But the occurrence of ESR further depends on the possibility of a spin-flip process. For this, the transport channel must be compatible with spin-flip processes.

\begin{figure}
\centering 
\includegraphics[width=1.1\linewidth]{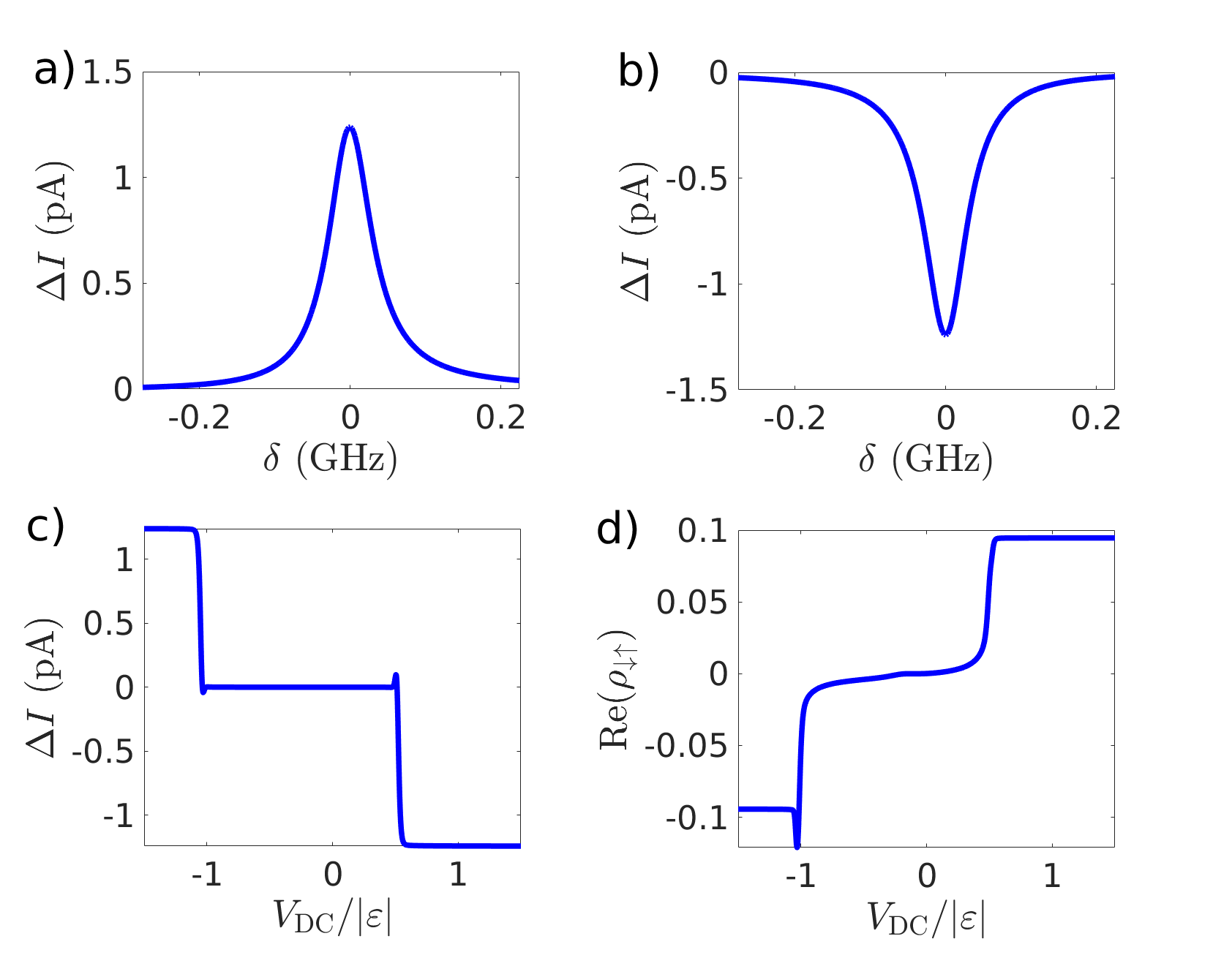}
\caption{ a) and b) ESR signal  $\Delta I(f)=I(f)-I_{BG}$ as function of relative frequency $\delta=f-f_0$ for two different signs of the DC bias. in a) the DC bias is negative and in b) the DC bias is positive which inverts the ESR amplitude. For this system the Larmor frequency is $f_0=17.025$ GHz which is the natural resonance frequency of the Hamiltonian plus the re-normalization imposed by the Lamb shift. c) ESR signal and d) real part of the coherence $\rho_{\uparrow \downarrow}$ between spin up and down as a function of DC bias when on resonance ($\delta=0$). The transport channels are closed for $V_{DC}\lesssim U+\varepsilon$ (neglecting the Zeeman energy) and $V_{DC}\geq \varepsilon$. In this work we took $U=3 |\varepsilon|/2$,
so the ESR signal is zero between $V_{DC}/|\varepsilon|\lesssim 0.5$ and $V_{DC}/|\varepsilon|\geq-1$. The behavior of the ESR signal reflects the behavior of the coherences except for a sign.} 
\label{IvsV}
\end{figure}

First, we study the dependence of the magnitude and sign of the ESR signal $\Delta I$ as function of the  magnitude and sign of the applied DC bias. Figure~\ref{IvsV} (a) and (b) show two representative spectra taken at opposite signs of the DC bias. The difference between both spectra is more than a change of sign. 
To better understand this behavior, Fig.~\ref{IvsV} (c) shows the ESR peak intensity as function of $V_{\rm{DC}}$. Take, for example, a positive bias where we obtain a large negative value of the ESR signal. This correlates with a large contribution of the coherence-term $\rho_{\uparrow \downarrow}$ between spin up and down (Fig.~\ref{IvsV} (d)). We emphasize that this occurs in the long-time limit under substantial decoherence of the system as long as the drive sustains the coherences. The connection between ESR signal and coherences of the density matrix can be understood by studying the behavior of the electronic current, Eq. (\ref{current_t}).

When the applied bias is positive ($\mu_L-\mu_R>0$), spin-polarized electrons flow from the left electrode into the impurity. A negative ion is formed if $\mu_L> \Delta_{2,\downarrow}=E_2-E_\downarrow\approx U+\epsilon$ (we have neglected the Zeeman energy), that corresponds to a transition from a singly-occupied level (with spin down, $u=\downarrow$) to a doubly charged level ($v=2$). At the same time, we need that $\mu_R<\Delta_{2,\downarrow}$, as is the case at positive bias. Similarly, the formation of the positively-charge ion is energetically possible. However, there is an important asymmetry due to the very different couplings between impurity and electrodes ($\gamma_L\ll\gamma_R$) as well as in the bias drop. As a consequence the formation of the negative ion is favored over the positive one for this present case.

Then, we can simplify the expression for the electron current, Eq. (\ref{current_t}), by neglecting the involvement of the positive ion, and only considering the negative ion as the intermediate step in the electron transfer between electrodes through the impurity:
\begin{eqnarray}
I(\omega)&=&
\frac{2e}{\hbar} \mbox{Re}  \bigg\{\rho_{\downarrow}(\omega) \Gamma_{\downarrow 2,2\downarrow,L;0}^{-} + \rho_{\uparrow}(\omega) \Gamma_{\uparrow 2,2\uparrow,L;0}^{-} + \nonumber \\
&&  \rho_{\downarrow,\uparrow}(\omega) \Gamma_{\downarrow 2,2\uparrow,L;-1}^{-}+\rho_{\uparrow,\downarrow}(\omega) \Gamma_{\uparrow 2,2\downarrow,L;1}^{-}
 \bigg\},
 \label{current_posV}
\end{eqnarray}
where, for instance, $\Gamma_{\downarrow 2,2\downarrow,L;0}^{-}$ is the electron rate for a process that involves a non-spin-flip transition (spin-up state) through the doubly-occupied one by exchanging an electron with the left electrode, Floquet index $n=0$. At the same time, $\rho_{\downarrow}=\rho_{\downarrow\downarrow,0}$ while $\rho_{\downarrow\uparrow}=\rho_{\downarrow\uparrow,1}$ and $\rho_{\uparrow\downarrow}=\rho_{\uparrow\downarrow,-1}$ where $-1,0,1$ are Floquet indices.

At a large-enough bias, all channels are open giving a background current, $I_{BG}$:
\[
I_{BG}=
\frac{2e}{\hbar} \mbox{Re}  \bigg\{\rho_{\downarrow}(\omega) \Gamma_{\downarrow 2,2\downarrow,L;0}^{-} + \rho_{\uparrow}(\omega) \Gamma_{\uparrow 2,2\uparrow,L;0}^{-} \bigg\},
\]
which recovers the usual expression for the current for very asymmetrical couplings~\cite{meir_wingreen_prl_1992}. The background current shows a small frequency dependence as it is largely given by the rates with Floquet index $n=0$. Indeed, there is no coherence in the density matrix when the driving frequency is different from the Larmor frequency (off resonance) and $I(\omega)=I_{BG}$. 

Only on resonance, is the coherence $\rho_{\downarrow,\uparrow}(\omega)$ different from zero. Then, there is a clear frequency-dependent contribution to the current at the Larmor frequency that originates in the coherences of the density matrix. Accordingly, the coherences contribution to the DC current depends on the Floquet indices $n=\pm 1$.

Increasing the value of the charging energy, $U$, moves the doubly-occupied state energy ($E_{2}=2\epsilon+U$). For $U\rightarrow+\infty$, it becomes impossible to open the channel connecting the single-electron states with the doubly-occupied one. As a consequence, the ESR signal completely disappears for positive bias.

At negative bias, $\mu_L<\Delta_{\downarrow,\emptyset}=-10$ meV marks the threshold for having a current, where $v=\emptyset$ corresponds to the positively charged impurity. As in the discussion above, we have neglected the Zeeman energy. The ESR signal also follows the behavior of $-\rho_{\uparrow \downarrow}$ as above, Fig. \ref{IvsV}. 

The intermediate state mediating the transport process at negative bias is the one corresponding to the positive ion, $v=\emptyset$. Then Eq. (\ref{current_t}) can be simplified by taking the positive ion contribution:
\begin{eqnarray}
I(\omega)&=&
-\frac{2e}{\hbar} \mbox{Re}  \bigg\{\rho_{\downarrow}(\omega) \Gamma_{\downarrow \emptyset,\emptyset\downarrow,L;0}^{+} + \rho_{\uparrow}(\omega) \Gamma_{\uparrow \emptyset,\emptyset\uparrow,L;0}^{+} + \nonumber \\
&&  \rho_{\downarrow,\uparrow}(\omega) \Gamma_{\downarrow \emptyset,\emptyset\uparrow,L;-1}^{+}+\rho_{\uparrow,\downarrow}(\omega) \Gamma_{\uparrow \emptyset,\emptyset\downarrow,L;1}^{+}
 \bigg\},
 \label{current_negV}
\end{eqnarray}
where again, the ESR signal originates in the coherences of the density matrix. Contrary to the positive-bias case, the limit $U\rightarrow\infty$ does not alter the results since the doubly-occupied level is not involved.

The presence of a finite charging energy then leads to breaking the electron-hole symmetry. At $U\rightarrow\infty$ the electron-hole asymmetry becomes the largest, with no ESR signal for positive bias and a large signal for negative bias at the bias threshold marked by the impurity level.

\subsection{ESR-STM linewidths}

\begin{figure}
\centering 
\includegraphics[width=1.05\linewidth]{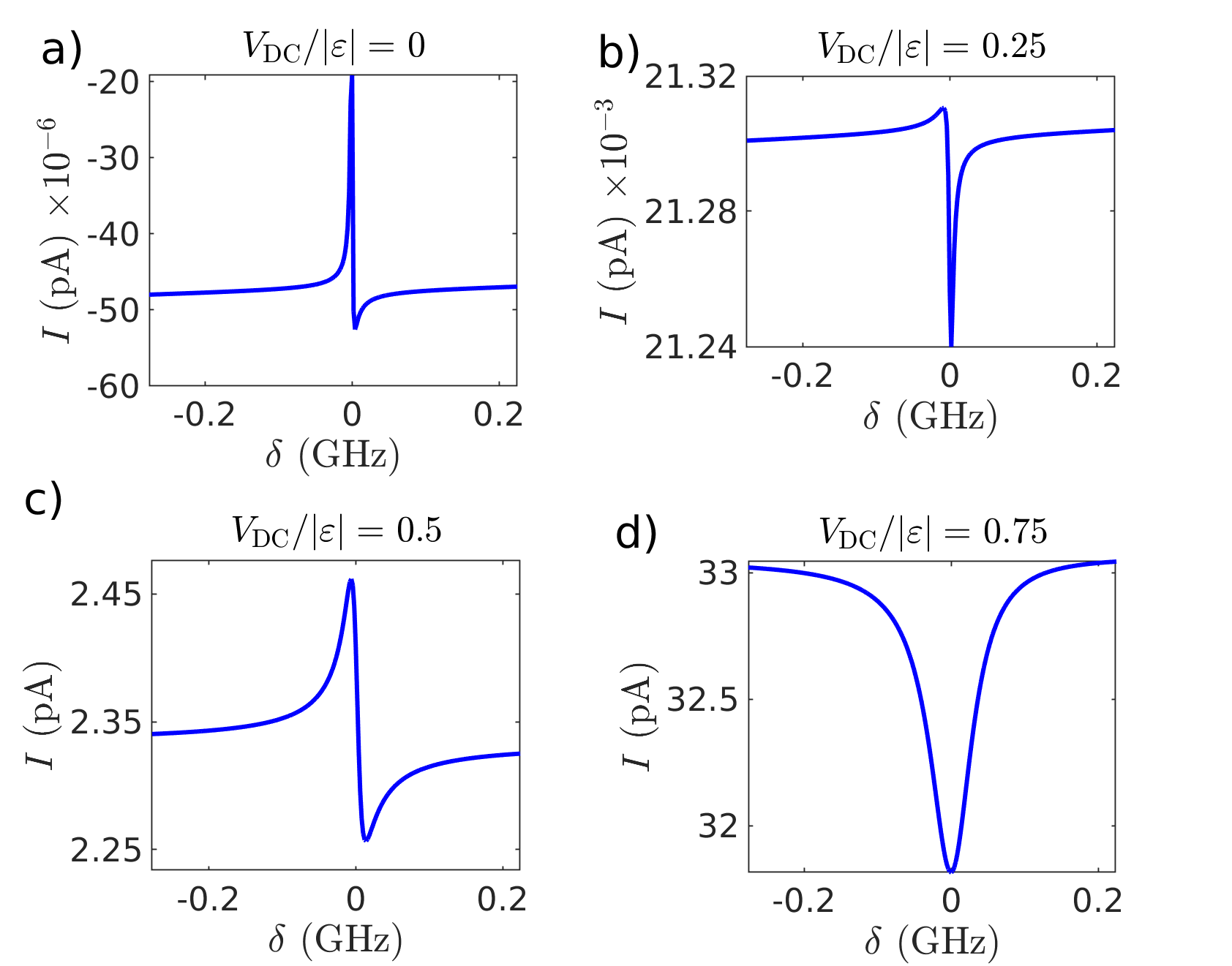}
\caption{DC current as a function of the driving frequency$\delta=f-f_0$, for the four different positive voltages a) $V_{DC}=0$, b) $V_{DC}=2.5$ mV, c) $V_{DC}=5$ mV and d) $V_{DC}=7.5$ mV. The background current was not removed. The current changes in a small interval about the resonance frequency. For DC bias below the threshold (at 5 mV here) the DC current drops dramatically as the channel closes and the line shape as a function of frequency becomes increasingly asymmetric. Moreover, the width of the resonance also increases with the DC bias, leading to smaller $T_2$ times as the decoherence is enhanced. The more asymmetric Fano profiles are found near the transport-channel thresholds.} 
\label{pos}
\end{figure}

Figure~\ref{pos} shows four characteristic CW ESR-STM signals as a function of the frequency of the drive, $f=\omega/2\pi$, for positive DC bias. At threshold, $V_{DC} \approx U+\varepsilon\approx 5$ mV, a strongly asymmetric Fano profile is obtained. This behavior can be traced back to the interference between the on-resonance scattering with the background. As the bias is further reduced, the transmission channel is increasingly closed, leading to a smaller background current and a smaller signal. In this regime, the ESR signal also depends on the change of the populations, in stark contrast to the open channel case, where the ESR signal is basically determined by the coherences.

This closed-channel region is of practical importance because here the system exhibits an enhanced coherence time.
The present treatment of this regime is valid as long as higher-order transport processes such as cotunneling are not dominating.

